\documentclass[useAMS,usenatbib]{mn2e}
\usepackage{natbib}
\usepackage{graphicx}
\usepackage{times}
\usepackage{rotate}
\usepackage{color}
\usepackage{ulem}
\newcommand{\simgt}{\lower.5ex\hbox{$\; \buildrel > \over \sim \;$}}
\newcommand{\simlt}{\lower.5ex\hbox{$\; \buildrel < \over \sim \;$}}
\newcommand{\ave}[1]{\left\langle #1\right\rangle}

\def\hMpc{h^{-1}{\rm Mpc}}

\begin{document}
\title[Testing subhalo abundance matching from redshift-space clustering]
{Testing subhalo abundance matching from redshift-space clustering}

\author[M. Yamamoto et al.]
{Mikito Yamamoto,$^1$ Shogo Masaki$^2$ and Chiaki Hikage$^3$  \\
$^1$ Department of Physics, Nagoya University,  Aichi 464-8602, Japan\\
$^2$ NTT Secure Platform Laboratories, NTT Corporation, Tokyo 180-8585, Japan\\
$^3$ Kobayashi Maskawa Institute (KMI), Nagoya University, Aichi 464-8602, Japan
}

\maketitle

\begin{abstract}
  We present a first application of the subhalo abundance matching
  (SHAM) method to describe the redshift-space clustering of galaxies
  including the non-linear redshift-space distortion, i.e., the
  Fingers-of-God. We find that the standard SHAM connecting the
  luminosity of galaxies to the maximum circular velocity of subhalos
  well reproduces the luminosity dependence of redshift-space
  clustering of galaxies in the Sloan Digital Sky Survey in a wide
  range of scales from $0.3$ to $40~\hMpc$. The result indicates that
  the SHAM approach is very promising for establishing a theoretical
  model of redshift-space galaxy clustering without additional
  parameters. We also test color abundance matching using two
  different proxies for colors: subhalo age and local dark matter
  density following the method by \cite{Masaki13a}. Observed
  clustering of red galaxies exhibits much stronger Fingers-of-God
  effect than blue galaxies.  We find that the subhalo age model
  describes the observed color-dependent redshift-space clustering
  much better than the local dark matter density model. The result
  infers that the age of subhalos is a key ingredient to determine the
  color of galaxies.
\end{abstract}


\section{Introduction}
\label{sec:intro}
Establishing connection between the properties of galaxies and the
underlying dark matter is crucial for both studies of galaxy evolution
and cosmology. Star formation histories of galaxies has been studied
by associating galaxies with their host dark matter halos and their
connection provides fundamental constraints on galaxy formation models
\citep[e.g.,][]{Conroy09,Leauthaud10,Behroozi13}. Future galaxy
surveys such as Prime Focus Spectrograph (PFS) \citep{PFS}, the Dark
Energy Spectroscopic Instrument (DESI) \citep{DESI}, {\it Euclid}
\citep{Euclid} and {\it the Wide Field Infrared Survey Telescope
  (WFIRST)} \citep{WFIRST} use both luminous red galaxies and emission
line galaxies to trace the large-scale structure at $z\simlt 2$. A
major uncertainty for the precision cosmology using galaxy surveys
comes from the challenge of relating galaxies and dark matter.

Subhalo abundance matching (SHAM) is a promising approach to relate
the properties of galaxies to dark matter subhalos
\citep[e.g.,][]{Kravtsov04,NagaiKravtsov05,Conroy06}. The simple
abundance matching model by assigning luminosity in the order of the
maximum circular velocity of dark matter subhalos successfully
reproduces the galaxy clustering at different redshifts
\citep{Conroy06}. \citet{Masaki13b} also finds a good abundance
matching between the progenitor halos of luminous red galaxies (LRGs)
and the massive halos at $z\sim 2$ and then explains the clustering
properties of LRGs very well. There also has been recent attempts to
relate the galaxy color to the subhalo properties, i.e., color
abundance matching. Galaxy color reflects the activity of on-going
star formation: red galaxies consists of aged stars and the star
formation is quenched, whereas blue galaxies are relatively young and
their star formation is active. It is also known that redder galaxies
live in denser environments via the measurement of galaxy clustering
\citep{Norberg02,Zehavi05,Coil08,Guo14,Skibba14} and also from the
color-density relation \citep{Balogh04,Cooper06,Blanton07,Bamford09}.
\citet{Masaki13a} extends a SHAM technique to explain color-dependent
properties of galaxy clustering as well as galaxy-galaxy lensing using
two proxies of color: one is the local dark matter density motivated
by the environmental dependence of galaxy color; the other is the
subhalo age reflecting the different aged population between red and
blue galaxies. \citet{Hearin13} also perform color abundance matching
by assigning the redshift $z_{\rm starve}$ characterizing the epoch of
star formation quenching to subhalos.

So far, the projected correlation function along the line-of-sight has
been commonly used for testing SHAM to avoid the effect of
redshift-space distortion (RSD) due to peculiar motion of galaxies.
The velocity of galaxies within and outskirts of clusters is
complicated and affected by different physics including the dynamical
friction, tidal stripping/disruption, merging, and ram pressure. The
internal motion of galaxies elongate the RSD of galaxies along the
line-of-sight direction, known as Fingers-of-God
\citep[FoG;][]{Jackson72}.  The FoG effect is clearly different by
colors: red galaxies show much stronger FoG effect than blue
galaxies \citep{Zehavi05,Coil08}. In the framework of the halo model,
the line-of-sight elongation in the galaxy distribution emerges from
the one-halo term, which is the contribution of central-satellite and
satellite-satellite galaxy pairs in the same halos
\citep{HikageYamamoto13}. The observed color difference of FoG feature
mainly reflects the difference of the satellite fraction, radial
profiles and their kinematics between red and blue galaxies.


In this paper we extend the SHAM approach to redshift space and test
if the model describes the luminosity and color dependence of
redshift-space galaxy clustering for the first time. We characterize
the anisotropy of redshift-space galaxy clustering by a multipole
expansion.  High-$l$ multipole components such as hexadecapole ($l=4$)
mainly generated from the FoG effect provide a useful probe of
constraining the fraction and the internal velocity dispersions of
satellite galaxies \citep{Hikage14}. In our analysis, we focus on the
non-linear scales from subMpc to $\sim 1$Mpc where the difference of
the FoG effect can be clearly seen. Extending SHAM to redshift space
is an important step for establishing a theoretical model of the
redshift-space clustering with a small number of parameters. The
coherent bulk motion of galaxies induced by gravitational evolution
squashes the distribution of galaxies along the line-of-sight
direction \citep{Kaiser87,Hamilton92,Peacock01}. The anisotropy in the
clustering provides a good probe of cosmic growth rate and have been
used for testing gravity
\citep[e.g,][]{Guzzo08,Yamamoto08,Yamamoto10,Beutler14}.  Very
recently \citet{Hearin15} points out the impact of the assembly bias
using color abundance matching technique and shows that the simple
formalism of halo occupation distribution causes a significant
systematics in RSD studies using pairwise velocity statistics.

The paper is summarized as follows: in Section 2, we explain the
details of the observational catalogs and subhalo samples used in our
analysis. We also make a brief summary of the two color assignment
schemes based on the subhalo age and local dark matter density. In
Section 3, we show the results of the luminosity and color dependence
of observed redshift-space clustering compared with those of subhalos.
and test SHAM and two color assignment methods. We
discuss what causes the difference of the two color assignment
methods. Section 4 is devoted to summary and conclusions.

\section{Observation and subhalo samples}
\label{sec:sample}
\subsection{Observational data}
\label{subsec:data}
We use the magnitude-limited samples of the Sloan Digital Sky Survey (SDSS)
Data Release 7 galaxies \citep{SDSSDR7} in three different
luminosity bins: $-22<M_r<-21$, $-21<M_r<-20$, and $-20<M_r<-19$
where $M_r$ is the r-band absolute magnitude including $-5\log h$
term. The apparent magnitude $m_r$ ranges from $14.5$ to $17.6$
based on the Petrosian magnitude and {\it K}-corrected to rest-frame
magnitude at $z=0.1$. The galaxy color is divided into red and blue on
the {\it K}-corrected $g-r$ color and $r$-band apparent magnitude plane: red
when $g-r > 0.21-0.03M_r$, otherwise blue \citep[see the details of
  the samples in][]{Zehavi11}. The fractions of red galaxies become
$0.63$, $0.53$, and $0.43$ from luminous to faint magnitude samples.  Due to
the physical size of the spectroscopic fiber, both spectra of adjacent
galaxies with the angular separation less than $55$ arcmin cannot be
measured simultaneously, i.e., ``fiber collision effect''. This
corresponds to the projected comoving scale $r_p=0.12~h^{-1}$Mpc at the
outer edge of the sample. In our analysis, we focus on the range of scales where
the fiber collision effect is unimportant.

\begin{table*}
\begin{center}
\caption{Basic information of subhalo catalogs for three magnitude
  samples: simulation box size $L_{\rm box}$, the output redshift of
  simulation snapshot $z_{\rm out}$, the number of subhalos $N_{\rm
    sub}$ matching to the observed number density, satellite fraction $f_{\rm sat}$
  of red and blue subhalo catalogs in two color assignment models:
  subhalo age (Age model) and local DM density (L.D. model).
\label{tab:subhalo}}
\begin{tabular}{cccccccccc}
\hline
\hline
& & & & \multicolumn{6}{c}{\raisebox{0ex}{$f_{\rm sat}$}} \\
\cline{5-10}
\raisebox{0ex}{$M_r$} & \raisebox{0ex}{$z_{\rm out}$} & \raisebox{0ex}{$L_{\rm box}$} & \raisebox{0ex}{$N_{\rm sub}$}
& & \multicolumn{2}{c}{\raisebox{0ex}{Age model}} & & \multicolumn{2}{c}{\raisebox{0ex}{L.D. model}}  \\
\cline{6-7}\cline{9-10}
& & & & \raisebox{0.5ex}{All} & red & blue & & red & blue \\
\hline
$[-22,-21]$ & $0.1$ & $300~h^{-1}$Mpc & $29969$ & $0.22$ & $0.28$ & $0.12$ & & $0.26$ & $0.16$ \\
$[-21,-20]$ & $0$   & $200~h^{-1}$Mpc & $43200$ & $0.32$ & $0.40$ & $0.21$ & & $0.42$ & $0.20$ \\
$[-20,-19]$ & $0$   & $200~h^{-1}$Mpc & $81600$ & $0.33$ & $0.42$ & $0.26$ & & $0.59$ & $0.13$ \\
\hline
\end{tabular}
\end{center}
\end{table*}

\subsection{Subhalo catalogs and color assignment}
\label{subsec:subhalo}
We use the subhalo samples corresponding to the magnitude ranges of
the observational samples. Below we summarize the catalogs briefly
\citep[see the details in][]{Masaki13a}. The subhalo catalogs are
constructed from $N$-body simulations using publicly available code
{\sc Gadget-2} \citep{Springel01a,Springel05}.  The initial condition
is a random-Gaussian field with the power spectrum based on a flat
$\Lambda$ cold dark matter model of the {\it Wilkinson Microwave
  Anisotropy Probe (WMAP)} 7-year results \citep{Komatsu11}:
$\Omega_{\rm m}=0.272, \Omega_{\rm b}=0.0441, \Omega_\Lambda=0.728,
H_0=100h=70.2~{\rm km}~{\rm s}^{-1}{\rm Mpc}^{-1}, \sigma_8=0.807$,
and $n_{\rm s}=0.961$. We employ 1024$^3$ dark matter (DM) particles
in each cubic simulation box. We use two simulation boxes with the side
length $L_{\rm box}$ and the redshift of output snapshot $z_{\rm out}$
depending on the range of magnitude as listed in Table \ref{tab:subhalo}.

Halos are identified using the Friends-of-Friends (FOF) algorithm with
the linking length of $0.2$ times the mean interparticle distance.
Satellite subhalos, dense clumps within each halo, are identified
using the {\sc SUBFIND} algorithm with the minimum number of DM
particles set to be $20$ \citep{Springel01b}. A central subhalo is
defined as the rest of DM particles without satellite subhalos (and
the `fuzz' component). The central positions of both central and
satellite subhalos are the density maximum rather than the center of
mass, and then the velocity of a subhalo is given by the mean velocity
of all inner particles. We neglect the inner velocity of central
galaxies against the halo bulk velocity.

We assume the monotonic relation between the galaxy luminosity
measured with $M_r$ and the maximum circular velocity $V_{\rm
  max}^{\rm acc}$ to match their abundance as $n_{\rm gal}(>L)=n_{\rm
  subhalo}(>V_{\rm max})$ \citep{Conroy06}. $V_{\rm max}^{\rm acc}$ is
the maximum value of the circular velocity of particles $V_{\rm
  circ}(R)=\sqrt{GM(<R)/R}$, where $R$ is the distance from the center
of each subhalo.  The maximum circular velocity for a central subhalo
is computed at the observed epoch, however, $V_{\rm max}$ for a
satellite is computed at the accretion epoch because subhalos inside
clusters lose their initial mass due to tidal stripping while the
stellar mass is tightly bound \citep{NagaiKravtsov05}.  Furthermore we
divide each magnitude sample into red and blue colors based on 
the following two models \citep{Masaki13a}.

\begin{enumerate}
\item Subhalo age model (``Age model'') \\ The galaxy color reflects the
  age of the stellar population: red galaxies consist of old stellar
  population when the star formation is quenched, whereas blue galaxies
  consist of younger stellar population with active star formation. We
  assign the color by the subhalo formation epoch $z_{\rm form}$,
  which is defined as the epoch when the maximum circular velocity of
  the subhalo first crosses some constant value $f$ times $V_{\rm
    max}^{\rm acc}$:
\begin{equation}
\label{eq:zform}
V_{\rm max}(z=z_{\rm form})=f\times V_{\rm max}^{\rm acc} ~~~~ (0<f<1),
\end{equation}

\item Local DM density model (``L.D. model'') \\ The other way of color
  assignment is based on the local DM density motivated by the
  observation that redder galaxies locate in denser environments
  \citep{Zehavi05,Mandelbaum06}. The local DM density $\rho(R_{\rm
    DM})$ of each subhalo is estimated from the number of DM particles
  within the distance of $R_{\rm DM}$ from the subhalo center.
\end{enumerate}

We assign color information with subhalos by assuming a monotonic
relation with $z_{\rm form}$, or $\rho (R_{\rm DM})$ to match the red
fraction of each observational sample.
Followed by \cite{Masaki13a}, a constant factor $f$ in the age model
is set to be $0.9$ for all magnitude bins to agree with the projected
clustering for both red and blue galaxies \citep{Zehavi11}.  The scale
$R_{\rm DM}$ in the L.D. model is set to be $250~h^{-1}$kpc for the
most luminous sample of $-22<M_r<-21$ and $200~h^{-1}$kpc for the
other samples.

The fraction of satellite galaxies play a significant role in the
small-scale clustering because of their large internal motion compared
to the central galaxies.  Table \ref{tab:subhalo} lists the satellite
fraction $f_{\rm sat}=N_{\rm sat}/(N_{\rm cen}+N_{\rm sat})$ of red
and blue samples in the two color assignment models. In both models,
red galaxy samples have a larger $f_{\rm sat}$ than the blue sample.
Luminous galaxy samples have smaller $f_{\rm sat}$, which reflect that
the luminous galaxy is more likely to be central galaxies rather than
satellites.  The difference of $f_{\rm sat}$ among the two color
assignment models is small expect for the faintest sample where
(smaller) satellite fraction of red (blue) galaxies in the L.D. model.

\begin{figure*}
\begin{center}
\includegraphics[width=14cm]{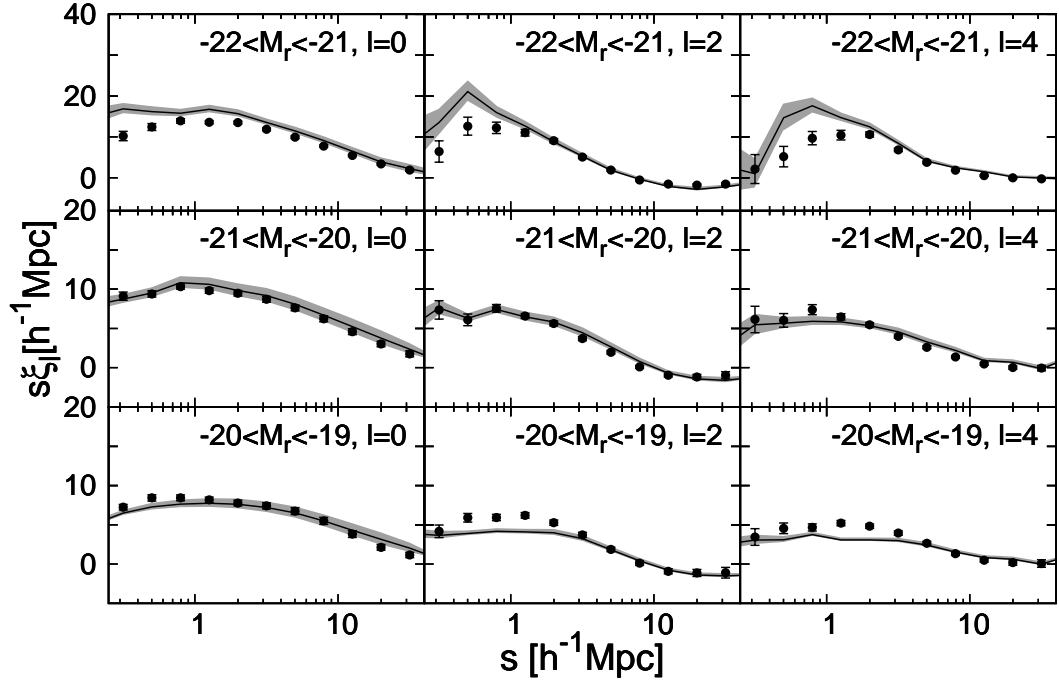}
\caption{Comparison of the multipole correlation functions $s\xi_l(s)$
  ($l=0,2,4$ from left to right) between observations (symbols) and
  subhalo catalogs (lines) for three magnitude samples. The error-bars
  represent the $1\sigma$ dispersion estimated with the Jackknife
  resampling method. Multiplication by the radial scale $s$ is for a
  visual purpose.}
\label{fig:nocolor}
\end{center}
\end{figure*}

\section{Results}
\label{sec:result}
In this section, we show the result of the redshift-space clustering
before and after color separation to see how SHAM and two color
assignment models reproduce the observations. We use multipole
correlation functions $\xi_l (l=0,2,4)$ and focus on the non-linear
gravitational scales from subMpc to $\sim10~$Mpc scales.

\subsection{Multipole correlation functions}
\label{subsec:xi}
We compute the redshift-space two-point correlation functions of
observational galaxy samples using Landy \& Szalay estimator
\citep{LandySzalay93}
\begin{equation}
\xi(s,\mu)=\frac{\rm DD-2DR+RR}{RR},
\end{equation}
where DD, DR, and RR is the number of data-data, data-random, and
random-random pairs normalized by the total number of pairs at each
bin of the three-dimensional distance of $s$ and the cosine of the
angle between the separation and the line-of-sight direction $\mu$.
Our bins of $s$ is logarithmically equal from $0.32$ to $40~h^{-1}$Mpc
with 11 bins.

The anisotropy of galaxy clustering in redshift space is described
with the multipole expansion using Legendre polynomials:
\begin{equation}
\xi_l(s)=\frac{1}{2}\int_{-1}^{1}d\mu \xi(s,\mu){\cal L}_l(\mu),
\end{equation}
where $\mu$ is the cosine of the angle to the line-of-sight direction
and ${\cal L}_l(\mu)$ is the $l$-th Legendre polynomials: ${\cal
  L}_0(\mu)=1$, ${\cal L}_2(\mu)=(3\mu^2-1)/2$, and ${\cal
  L}_4(\mu)=(35\mu^4-30\mu^2+3)/8$. 

In our analysis, we focus on the monopole $(l=0)$, quadrupole $(l=2)$,
and hexadecapole $(l=4)$ components. Coherent bulk motion of galaxies
known as Kaiser effect squashes the redshift-space galaxy clustering
along the line-of-sight direction and then the quadrupole component
becomes negative.  Kaiser effect on higher-order multipoles is small.
On the other hand, the elongated distribution by the FoG effect
generates high-order anisotropic components such as a hexadecapole and
then both quadrupole and hexadecapole components have positive values
\citep{HikageYamamoto13}. As the FoG effect increases, their amplitude
becomes larger.

Since both observational and simulated subhalo samples are one
realization in each magnitude bin, we estimate the error of $\xi_l$
using the Jackknife re-sampling method as follows:
\begin{equation}
\sigma_l^2(s)=\frac{N-1}{N}\sum_i^{N} \left[\xi_l^i(s)-\ave{\xi_l(s)}\right]^2,
\end{equation}
where $N$ is the number of subsamples, $\xi_l^i$ denotes $l$-th
multipole for the sample without data in $i$-th sub-volume, and
$\langle\xi_l(s)\rangle$ represents the ensemble average of
$\xi_l^i(s)$ over subsamples.  In order to estimate the error, we
divide each observational sample into $125$ subsamples with equal sky
area. Subhalo samples are also divided into $125 (= 5^3)$ sub-cubes
with equal volume.

\begin{figure*}
\begin{center}
\includegraphics[width=14cm]{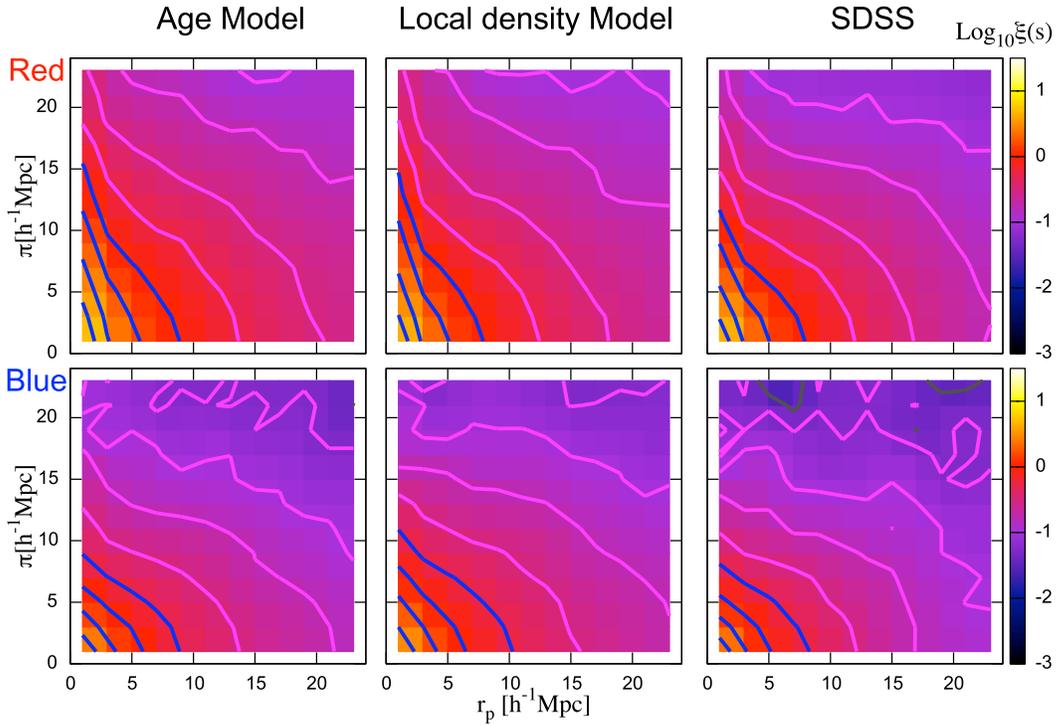}
\caption{Color difference of the two dimensional correlation function
  $\xi(r_p,\pi)$ where $r_p$ is the projected separation and $\pi$ is
  the line-of-sight separation.  Upper (Lower) panels show the results
  of red (blue) subhalos by separating color by the subhalo age (left)
  or the local DM density (center). Right panels show the
  observational results for comparison. All figures correspond to the
  sample with the intermediate magnitude range of $-21<M_r<-20$.}
\label{fig:xi2d}
\end{center}
\end{figure*}

\begin{figure*}
\begin{center}
\includegraphics[width=14cm]{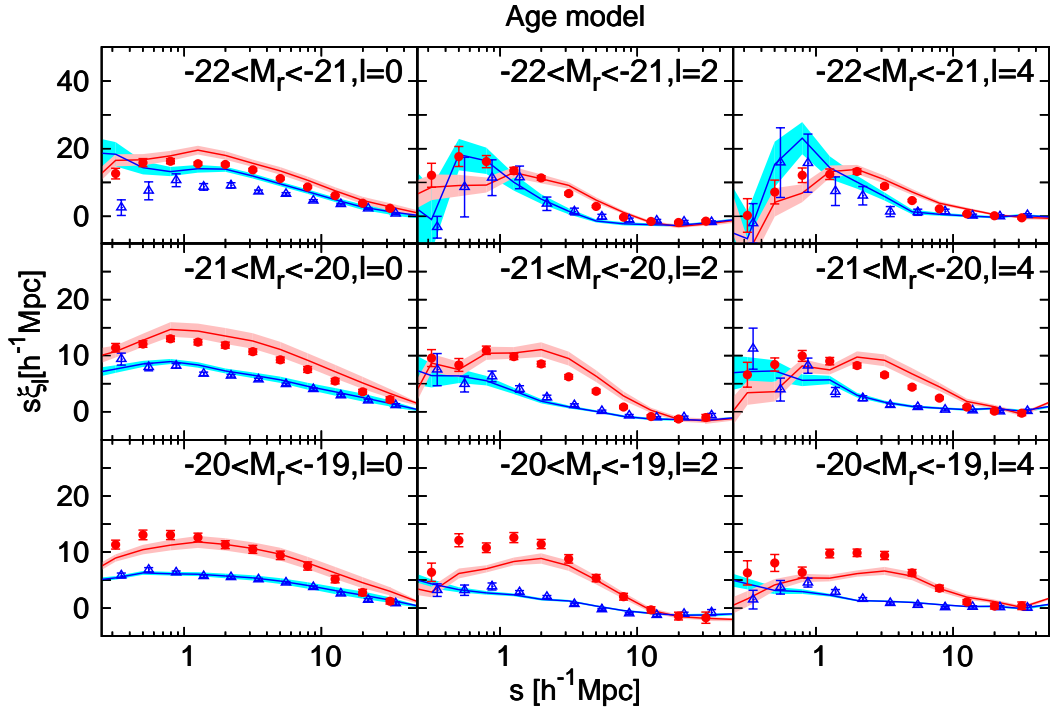}
\caption{Comparison of the observed multipole correlation functions
  $\xi_l(s)$ ($l=0,2,4$) for red and blue galaxies with those of
  simulated subhalo samples where the color is assigned by the subhalo
  age. Observational results for red and blue galaxies are plotted
  with filled circles and open triangles respectively. The simulation
  results are represented by the shade regions. Both errors of the
  observations and the simulations are estimated with the Jackknife
  resampling method.
\label{fig:xil_zform}}
\end{center}
\end{figure*}

\begin{figure*}
\begin{center}
\includegraphics[width=14cm]{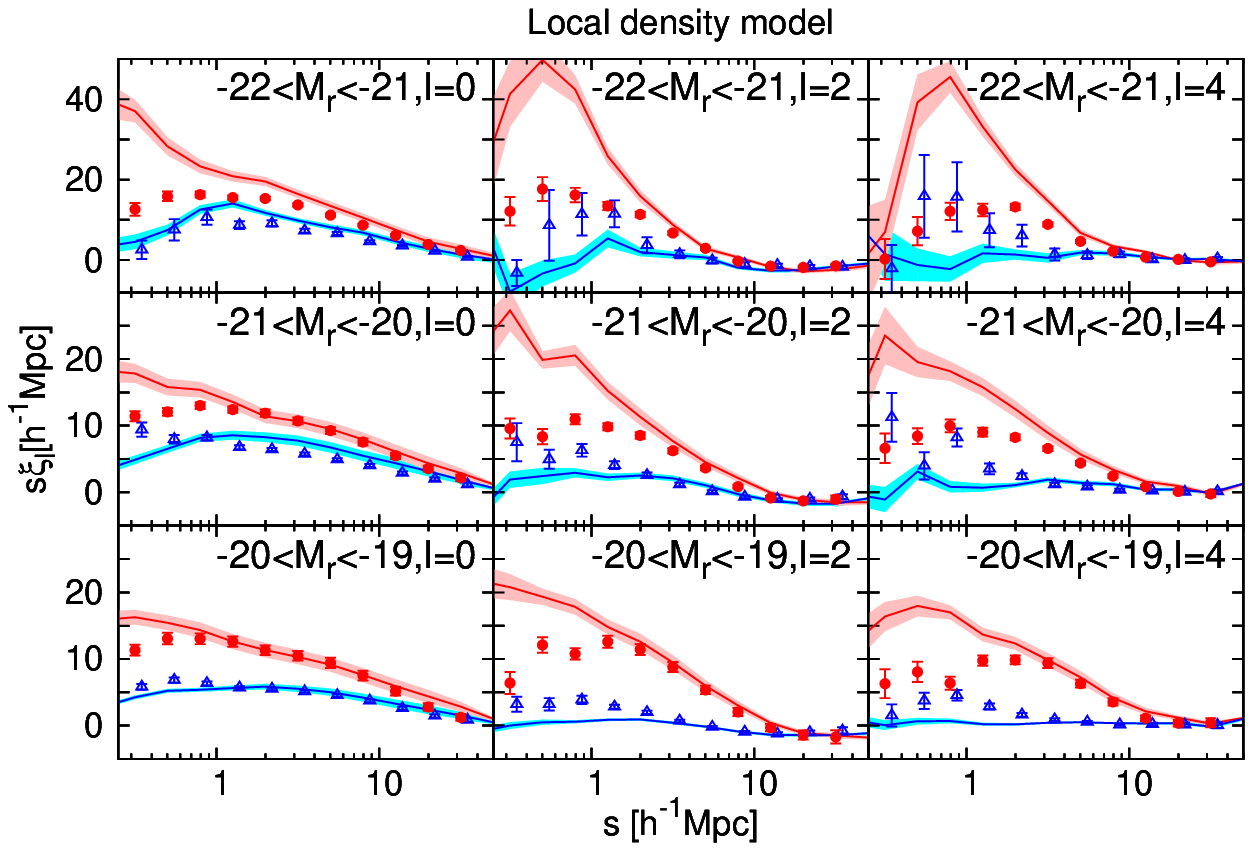}
\caption{Same as Fig. \ref{fig:xil_zform} but for the color
  assignment based on the local DM density model.
\label{fig:xil_rho}}
\end{center}
\end{figure*}

\subsection{Luminosity dependence}
\label{subsec:luminosity}
First we show the comparison of redshift-space clustering before color
separation between the observed galaxies and the SHAM-based subhalo
samples. Fig.  \ref{fig:nocolor} shows the results of $\xi_l(s) (l=0,
2, 4)$ for three magnitude samples. The errors of both the
observations and the subhalo samples are estimated from the Jackknife
resampling method.  We find that the SHAM method well reproduces
  the luminosity dependence of redshift-space galaxy clustering
  without introducing additional parameters. The overall amplitude of
  $\xi_l$ increases for more luminous samples because luminous
  galaxies are hosted by massive halos and has a larger galaxy
  biasing. One can find that $\xi_2$ and $\xi_4$ are suppressed on
  subMpc scales in the most luminous sample. This is because more
  luminous galaxies have a lower $f_{\rm sat}$ (see Table
  \ref{tab:subhalo}) and thereby the FoG anisotropy due to satellites
  becomes smaller. The agreement of $\xi_l$ between the observations
  and the subhalos is particularly excellent for the intermediate
  magnitude sample.  This indicates that the internal motion of
  satellites responsible for the FoG effect is well described by the
  motion of subhalos and then the SHAM approach can be a powerful tool
  for building a theoretical framework of redshift-space
  clustering. There is however a systematic difference that the
  brightest (faintest) galaxy sample has weaker (stronger) FoG effect
  than the subhalo sample. This may indicate the importance of
  baryonic physics such as hydrodynamic drag by the gas that is not
  included in $N$-body simulations. Hydrodynamical simulations show
  that baryonic components resists the tidal disruption of
  slowly-rotating subhalos and then decreases the averaged velocity of
  satellite subhalos \citep{Faltenbacher05,Wu13}. Our results may
  depend on the algorithms of finding subhalos. It would be
  interesting to apply other halo finders \citep[e.q.,][]{Behroozi13},
  however, these works are beyond the scope of our paper and we leave
  them to future works.

\subsection{Color dependence: age model vs local DM density model}
\label{subsec:color}
Next we see the color dependence of redshift-space clustering and test
two color assignment models based on the subhalo age and the local DM
density. Fig. \ref{fig:xi2d} shows the two-dimensional correlation
functions $\xi(r_p,\pi)$ where the distance of galaxy pairs are
separated into the tangential separation $r_p$ and the line-of-sight
separation $\pi$. Upper (lower) panels show the results of red (blue)
galaxies in the age model (left), the L.D. model (center), and
observations (right) for the intermediate magnitude sample.  Kaiser
effect due to the coherent infalling motion squash the galaxy
distribution along the line-of-sight direction on large scale.  On
small scale around 1$h^{-1}$Mpc, the FoG effect due to the internal
motion of galaxies inside clusters elongate the distribution of
galaxies along the line-of-sight direction. One can clearly see the
color difference of the FoG effect in both color assigning models: red
galaxies has much stronger FoG effect than blue galaxies. This feature
is consistent with the observational results for the SDSS galaxy
samples in the right panels and also with the previous work done by
\cite{Zehavi11}. One of the reason is that red galaxy samples have a
larger satellite fraction than blue ones as shown in Table
\ref{tab:subhalo}. The internal velocity dispersion of red galaxies is
also larger because red galaxies are hosted by massive halos. We make
further discussion in the next subsection \ref{subsec:discussion}.

We make a more detailed comparison using a multipole expansion.
Fig. \ref{fig:xil_zform} shows the measurements of monopole ($l=0$),
quadrupole ($l=2$), and hexadecapole ($l=4$) components for red and
blue galaxies of the three magnitude samples. We find that the red
galaxies have larger amplitudes than blue galaxies for all of the
multipoles in our focused range of scales where the FoG effect is
important.  For comparison, we plot the results of subhalo samples
with color assigned by the subhalo age. We find that the age model
reproduces the color difference of redshift-space clustering down to
subMpc scales very well. There are some systematic deviations: $\xi_l$
for blue subhalos in the most luminous bin are stronger, but $\xi_l$
for red subhalos in the faintest bin are weaker. The systematic trend
is also seen before color separation (see Fig. \ref{fig:nocolor} for
comparison). This indicates that the discrepancy is mainly due to the
incompleteness of SHAM not the color assignment scheme based on the
subhalo age.  We also compare the observations with the L.D. model in
Fig. \ref{fig:xil_rho}. In contrast to the age model, the L.D. model
has significant deviations for all of the magnitude samples and
multipoles at less than a few Mpc scale: red (blue) subhalos have too
large (small) amplitude compared to the observations. This means that
the color assignment by the local DM density is not enough to describe
the redshift-space clustering and that the subhalo age is a much
better proxy of galaxy color.

We evaluate the agreement of the multipole correlation functions
between the observations and the subhalo samples in the chi-squared
basis. The total $\chi^2$ in each magnitude bin is computed by summing
up the $\chi^2$ for red and blue galaxies over different multipoles
$l=0, 2,$ and 4:
\begin{equation}
\chi^2 = \sum_{\rm color}\sum_l^{0,2,4}
\sum_i^{s_{\rm min}<s_i<s_{\rm max}}\frac{[\xi_l^{\rm obs}(s_i)-\xi_l^{\rm model}(s_i)]^2}{\sigma_l^{\rm obs,~2}(s_i)+\sigma_l^{\rm model,~2}(s_i)},
\end{equation}
where $\sigma_l^{\rm obs,~2}$, and $\sigma_l^{\rm model,~2}$ are the
variance of $\xi_l$ for observations and subhalo samples estimated
from the Jackknife method. The range of the fitting scale is from
$s_{\rm min}=0.32h^{-1}$Mpc to $s_{\rm max}=40h^{-1}$Mpc with 11 bins
. We neglect the covariance between different scales, multipoles and
colors for simplicity.

\begin{figure}
\begin{center}
\includegraphics[width=8cm]{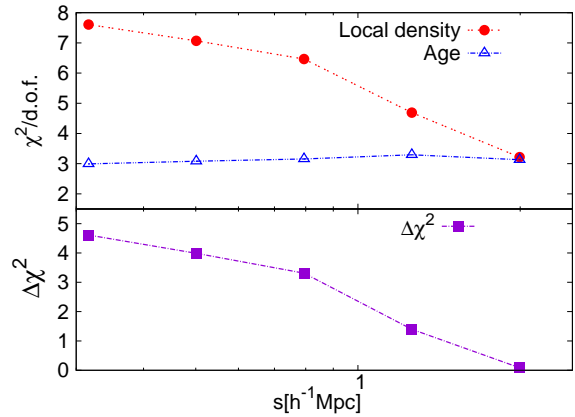}
\caption{Comparison of the reduced chi-squared values between the age
  model and the L.D. model on different bins of scale $s$.  Lower
  panel shows the difference of the chi-square $\Delta \chi^2$ between
  the two models.
\label{fig:chi2}}
\end{center}
\end{figure}

Table \ref{tab:chi2} lists the reduced chi-squared values of three
magnitude samples before and after color separation. The
degrees-of-freedom (d.o.f.) of each magnitude bin is $33$ (11 bins of
scale $\times$ 3 multipoles) before color separation and $66$ after
color separation. We find that the reduced $\chi^2$ of the age model
is $2$-$3$.  On the contrary, the L.D. model is significantly larger
chi-squared values such as $6$-$8$.  Fig. \ref{fig:chi2} shows the
comparison of the reduced $\chi^2$ at each bin of scale on $\simlt
1h^{-1}$Mpc between the two color models. The fitting of the
L.D. model is already worse around 1$h^{-1}$Mpc and then the deviation
increases as the scale goes down.

\begin{table}
  \begin{center}
    \caption{Reduced chi-squared values of the multipole power spectra
      $\xi_l (l=0,2,4)$ for three magnitude samples before color
      separation (``No color sep.'') and two color models (Age and
      L.D. model).  Each chi-squared value is summed over three
      multipoles $l=0, 2$, and $4$ and two colors after color
      separation. We adopt the range of scales is from
      $0.32~h^{-1}$Mpc to $40~h^{-1}$Mpc and the number of bins are 11
      in the range.  The total d.o.f. is $33$ (or $66$) before (or
      after) color separation.
      \label{tab:chi2}}
    \begin{tabular}{cccc}
      \hline
      \hline
      & \multicolumn{3}{c}{\raisebox{0ex}{$\chi^2$/d.o.f.}} \\
      \cline{2-4}
      \raisebox{1ex}{$M_r$} & \raisebox{0ex}{No color sep.} & \raisebox{0ex}{Age model} & \raisebox{0ex}{L.D. model} \\
      \hline
      $[-22,-21]$ & $3.49$ & $3.52$ & $8.45$ \\
      $[-21,-20]$ & $0.89$ & $1.87$ & $6.11$ \\
      $[-20,-19]$ & $4.01$ & $2.77$ & $6.18$ \\
      \hline
    \end{tabular}
  \end{center}
\end{table}

\begin{figure*}
\begin{center}
\includegraphics[width=8cm]{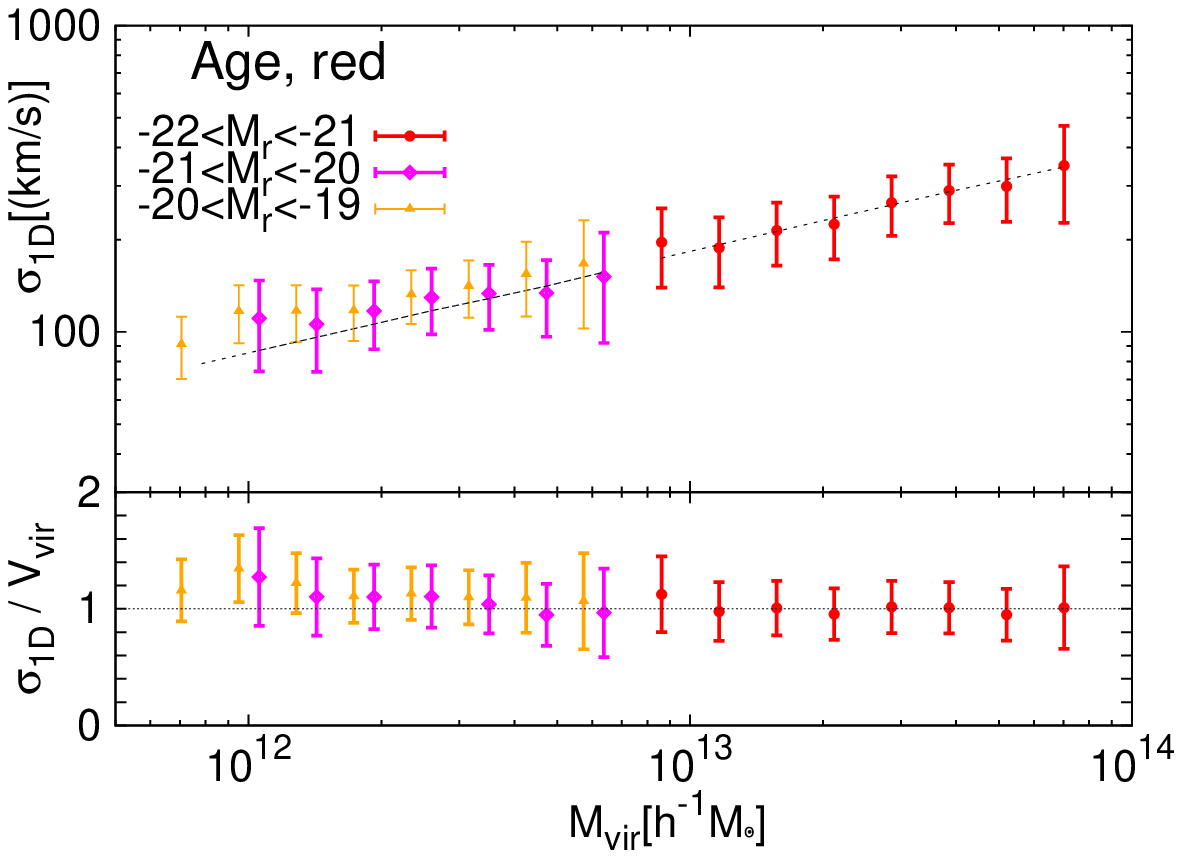}
\includegraphics[width=8cm]{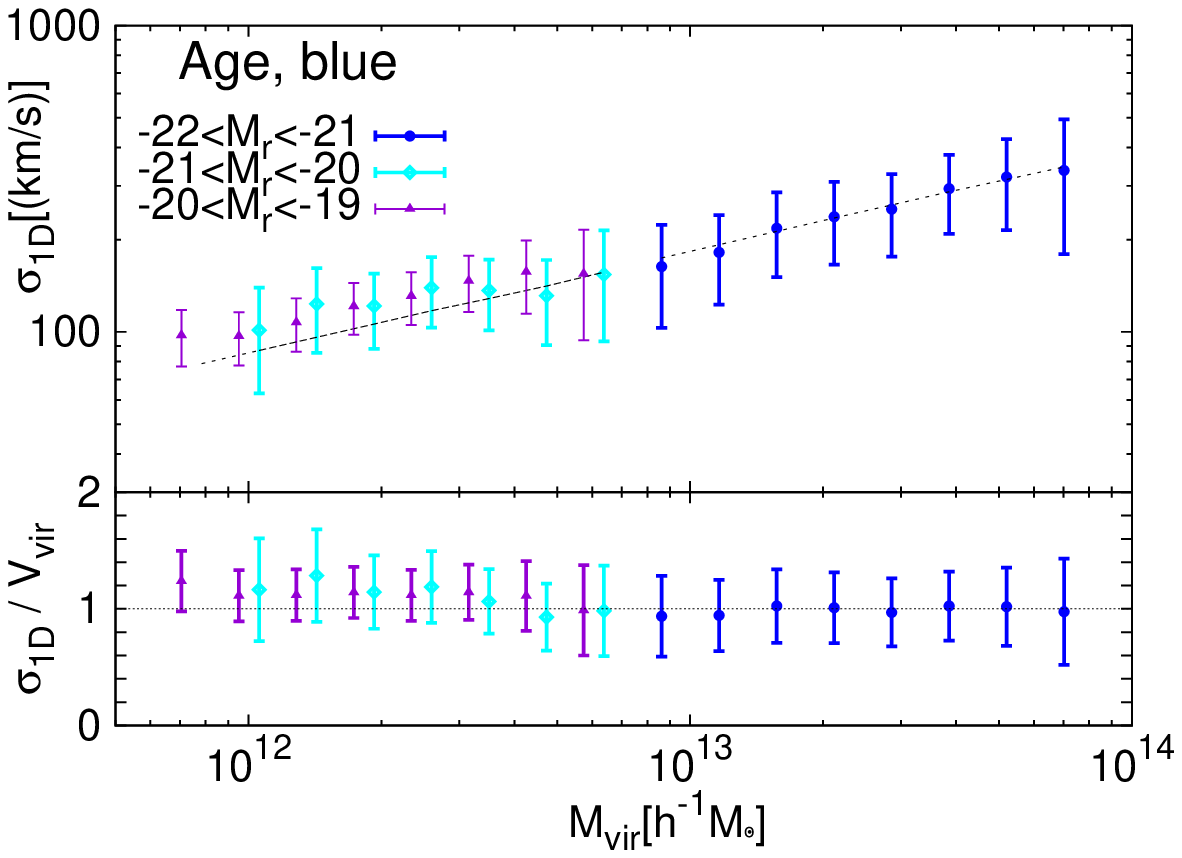}
\includegraphics[width=8cm]{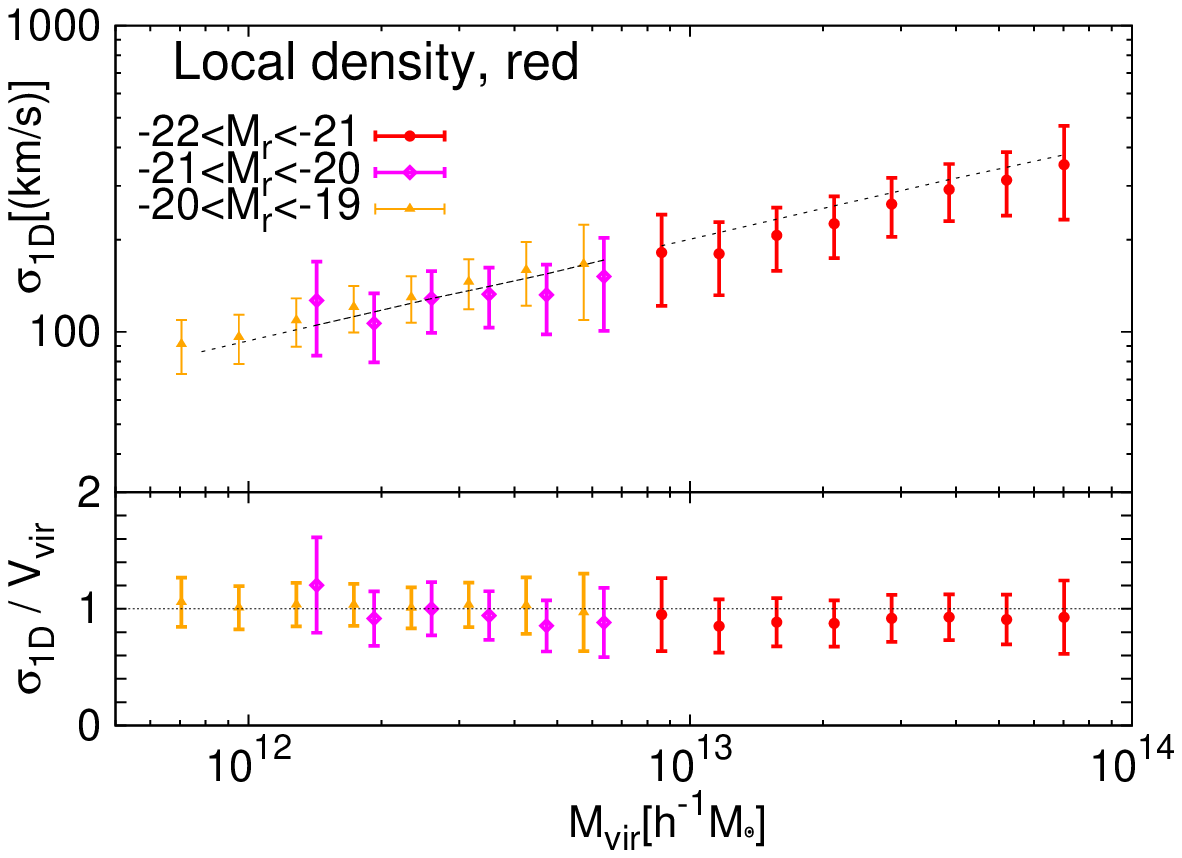}
\includegraphics[width=8cm]{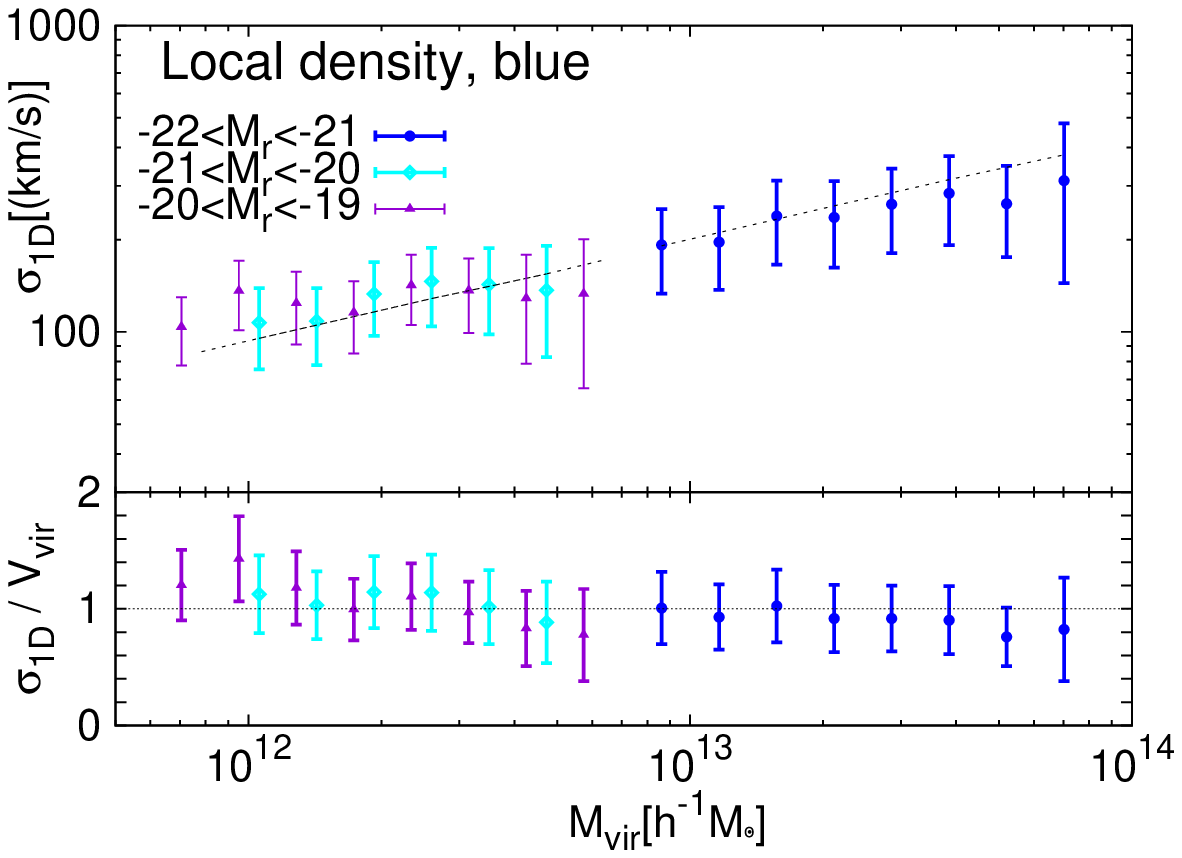}
\caption{One-dimensional internal velocity dispersion $\sigma_{\rm
    1D}$ of red (left) and blue (right) satellite subhalos relative to
  the central subhalos as a function of the host halo mass for three
  magnitude samples. The color separation is based on the age (upper)
  and local DM density (lower) models respectively.  For comparison,
  Virial velocity dispersion is plotted with lines. The error-bar
  represents 1$\sigma$ dispersion in each bin of halo mass. The lower
  panels show the ratio $\sigma_{\rm 1D}/\sigma_{\rm vir}$.
\label{fig:sigv}}
\end{center}
\end{figure*}

\subsection{Discussion}
\label{subsec:discussion}
In the previous subsection, we find that the age model has a better
agreement with the observed redshift-space clustering than the L.D.
model. We here investigate why the two models have different features
in the redshift-space clustering. The amplitude of $\xi_l$ from subMpc
to Mpc scale is mainly affected by the FoG effect, which depends on
the fraction of satellite galaxies, the internal velocity dispersion,
and also the radial profile of satellites in their host halos. We here
investigate where the differences between the models come from one by
one.

As the satellite fraction increases, the contribution of
central-satellite pairs and satellite-satellite pairs in the same
halos increases, which generate the large amplitude of quadrupole and
hexadecapole components on $\simlt 1$Mpc. As shown in Table
\ref{tab:subhalo}, however, the satellite fraction between the two
color models is roughly same except for the faintest magnitude bin and
thereby the number of satellites is not a main reason for the model
difference of redshift-space clustering.

The FoG effect comes from the internal velocity dispersion, which is
mainly dependent on the host halo mass. Table \ref{tab:hosthalo} lists
the mean halo mass weighted with the number of satellite subhalos:
\begin{equation}
\label{eq:mhalo_sat}
\ave{M_{\rm halo}^{\rm sat,wei}}=\frac{\sum_i M_{\rm halo,i}N_{\rm sat,i}}{\sum_i N_{\rm sat,i}},
\end{equation}
where $M_{\rm halo,i}$ and $N_{\rm sat,i}$ is the mass and the number
of satellite subhalos in $i$-th halo. One can see that more luminous
and redder satellites are hosted by more massive halos. We find that
the difference of the host halo mass between colors is rather small in
the L.D. model than the age model. It is thereby difficult to explain
the large color difference of redshift-space clustering in the
L.D. model.

We also see if the velocity dispersion of red and blue subhalos follow
the expectations of Virial theorem.  Fig. \ref{fig:sigv} shows the
one-dimensional (1D) internal velocity dispersion of satellite
subhalos as a function of the host halo mass in the age and
L.D. model. Lines represent the Virial velocity dispersion estimated
by $\sigma_{\rm vir}=(GM_{\rm vir}/2R_{\rm vir})^{1/2}$ where $M_{\rm
  vir}$ and $R_{\rm vir}$ represent Virial mass and radius. The 1D
velocity dispersion of satellite subhalos to the central subhalos are
estimated as $|\mathbf{v}^{\rm (sat)}-\mathbf{v}^{\rm
  (cen)}|^2/3$. The error-bars represent the 1-$\sigma$ dispersion in
each bin of halo mass. In both models, we find that the satellite
velocity dispersions agree with the expectations from Virial theorem
within the error bars irrespective of color, though dispersion
slightly decreases as the host halo is more massive.  The velocity
dispersion is unlikely to explain the difference of redshift-space
clustering between the two models.
\begin{table}
\begin{center}
\caption{Comparison of the mean halo mass weighted with the number of
  satellite subhalos (eq.[\ref{eq:mhalo_sat}]) for the red and blue
  subhalo samples in the two color assignment models.
\label{tab:hosthalo}}
\begin{tabular}{cccccc}
\hline
\hline
& \multicolumn{5}{c}{\raisebox{0ex}{$\ave{M_{\rm halo}^{\rm sat,wei}}$ [$10^{14}h^{-1}M_\odot$]}} \\
\cline{2-6}
\raisebox{0ex}{$M_r$} & \multicolumn{2}{c}{\raisebox{0ex}{Age model}} & & \multicolumn{2}{c}{\raisebox{0ex}{L.D. model}} \\
\cline{2-3}\cline{5-6}
& red & blue & & red & blue \\
\hline
$[-22,-21]$ & $1.46$ & $1.09$ & & $1.40$ & $1.36$ \\
$[-21,-20]$ & $1.30$ & $0.80$ & & $1.13$ & $1.16$ \\
$[-20,-19]$ & $1.06$ & $0.64$ & & $0.90$ & $0.76$ \\
\hline
\end{tabular}
\end{center}
\end{table}

\begin{figure*}
\begin{center}
\includegraphics[width=5cm]{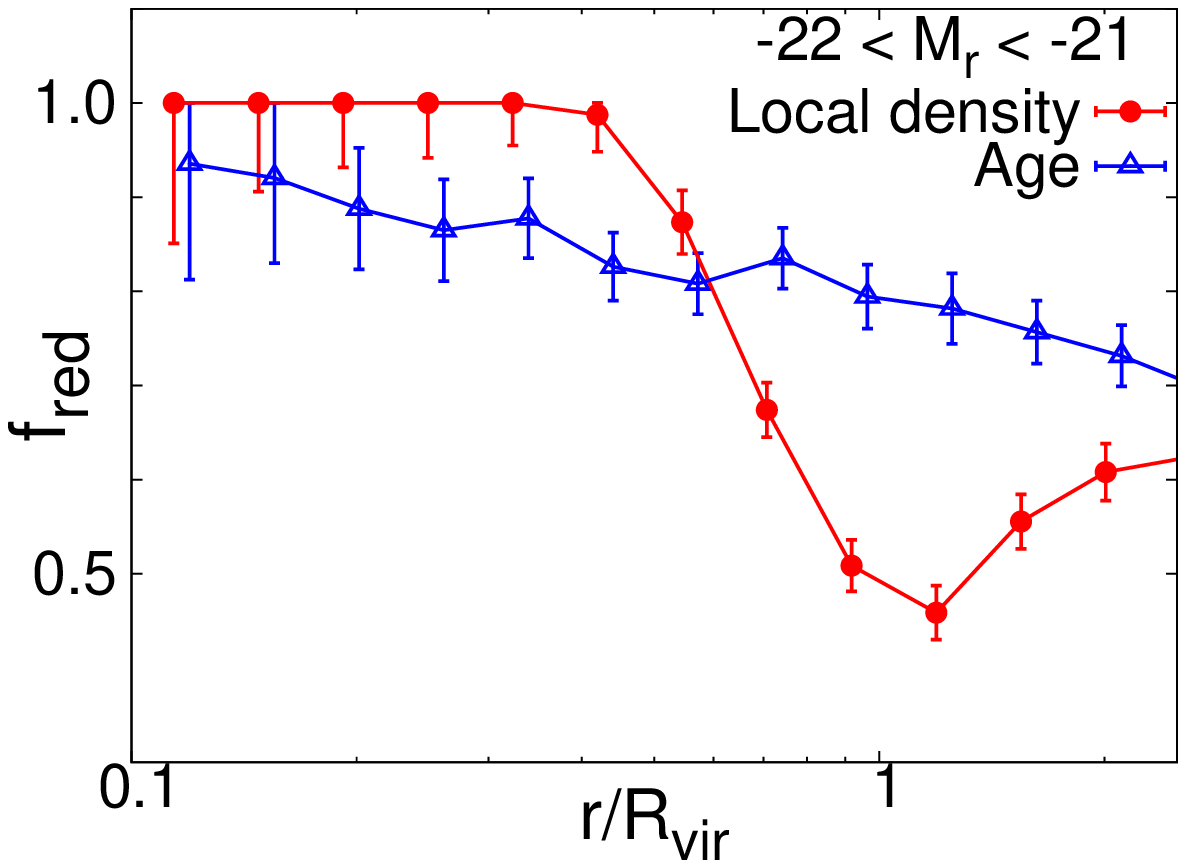}
\includegraphics[width=5cm]{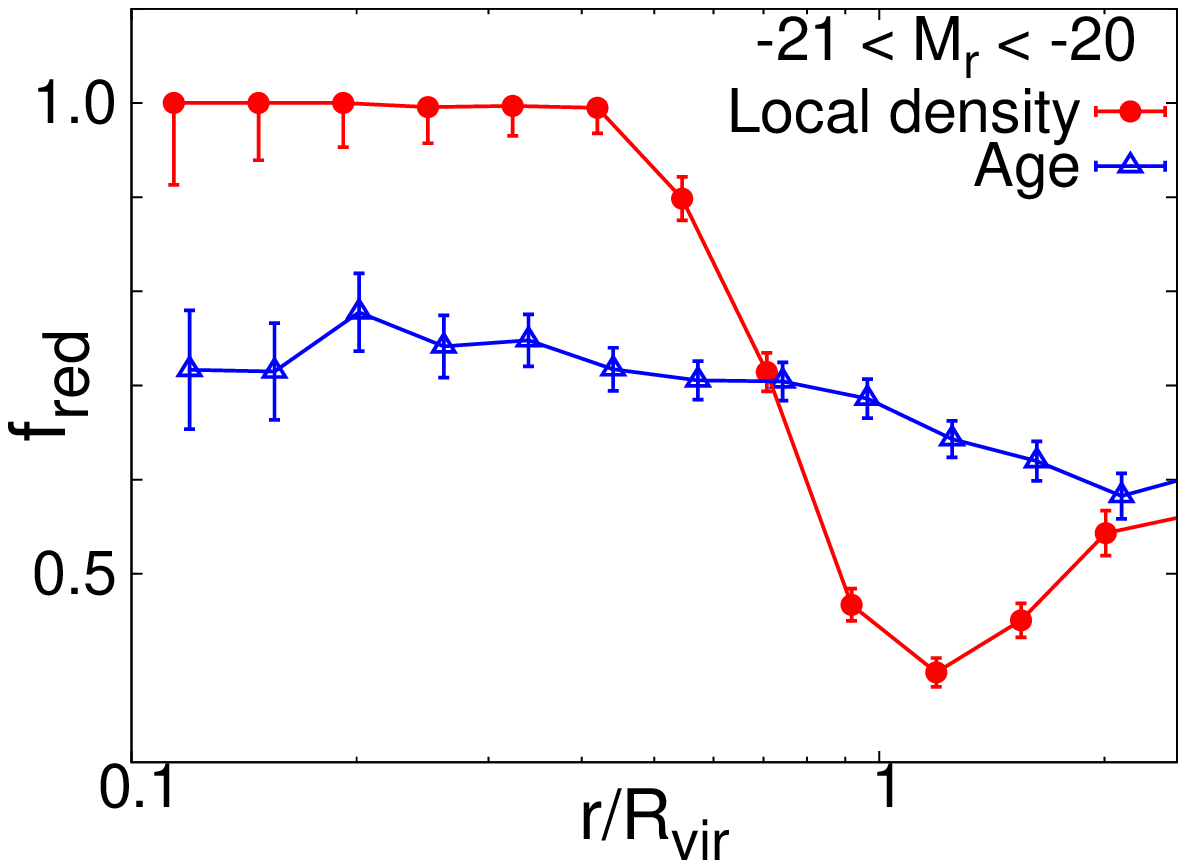}
\includegraphics[width=5cm]{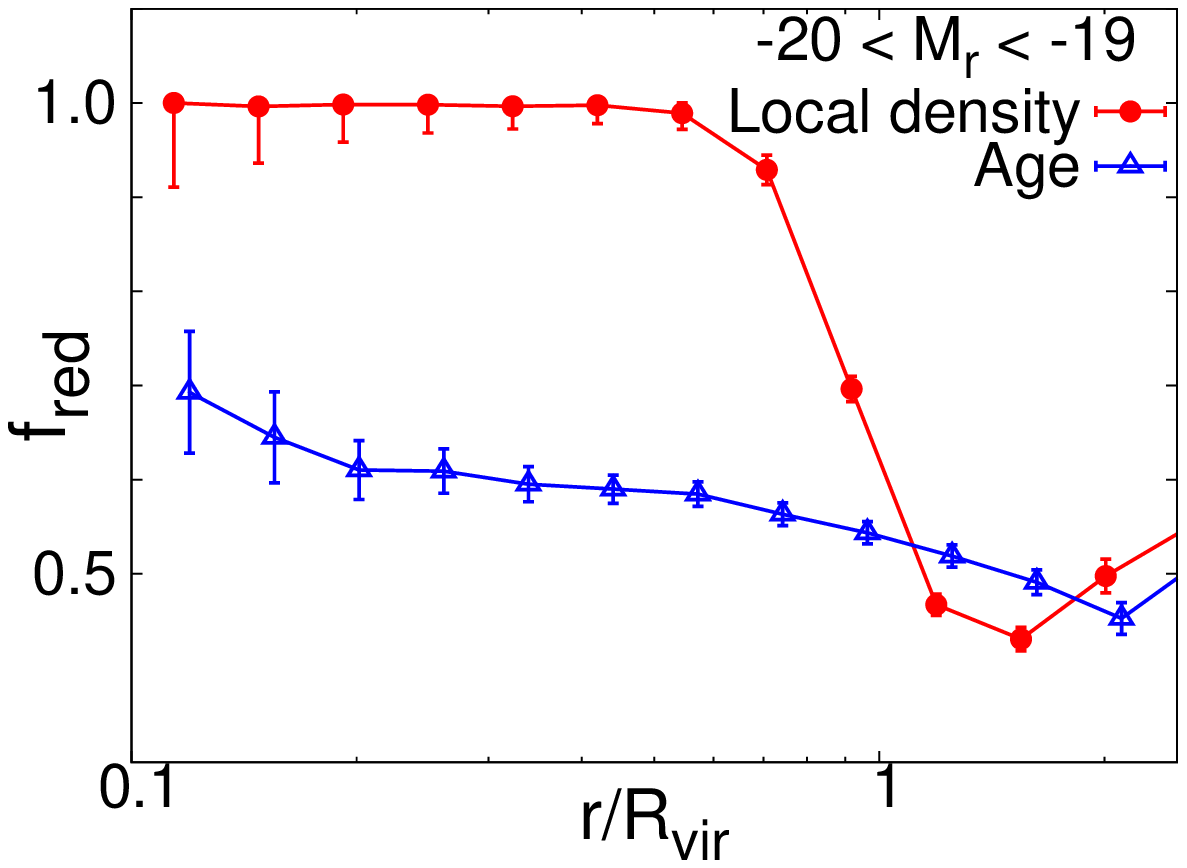}
\caption{The fraction of red galaxies as a function of the distance
  from the host halo center normalized by the Virial radius $R_{\rm
    vir}$ for three magnitude samples. Red and blue lines represent
  the results for the L.D. model and the age model respectively. The
  error-bar represents the Poisson error in each bin of scale.
\label{fig:redfrac}}
\end{center}
\end{figure*}

Finally we see the radial profile of satellites which affect the
small-scale clustering.  Fig. \ref{fig:redfrac} shows that the
fraction of red galaxies $f_{\rm red}\equiv N_{\rm red}/(N_{\rm
  red}+N_{\rm blue})$ as a function of the distance $R$ from the halo
center normalized by the Virial radius $R_{\rm vir}$ in the three
magnitude samples.  For all of the samples, there is a clear
difference between the two color models.  The red fraction in the age
model slightly decreases at larger $R$, whereas the red fraction in
the L.D. model is almost unity inside the host halo and drastically
declines around the Virial radius. This can explain the big
  difference between colors in the redshift-space clustering: a large
  number of red galaxy pairs inside halos causes too strong FoG
  effect, however, the lack of blue galaxies have too weak FoG.  The
observed red fraction actually shows a similar trend to the age model
\citep{Hansen09}. This means that the color assignment in the
L.D. model is too simple to describe the color difference of galaxy
clustering inside halos.

\begin{figure*}
\begin{center}
\includegraphics[width=14cm]{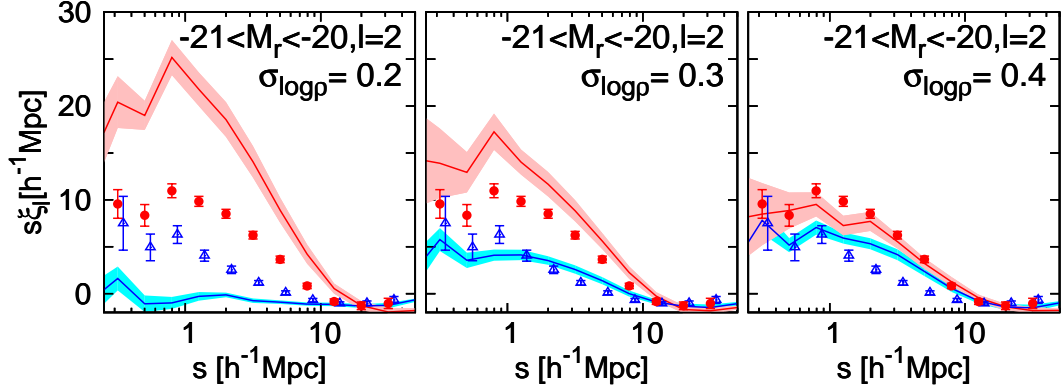}
\caption{Comparison of observed quadrupole component $\xi_2$ for red
  and blue galaxies based on the L.D. model. The scatter in the color
  assignment is included as the equation \ref{eq:scatter} and the
  logarithmic scatter value $\sigma_{\log\rho}$ is $0.2$, $0.3$, and $0.4$ from
  left to right panels.
\label{fig:xil_rho_scatter}}
\end{center}
\end{figure*}

We investigate how the result changes by including scatter in the
color assignment.  Instead of assuming the monotonic relation between
the color and $\rho_{\rm DM}$, we assign red color to the subhalo with
the local DM density of $\rho_{\rm DM}$ in the following probability:
\begin{equation}
\label{eq:scatter}
P_{\rm red}(\rho_{\rm DM})=\frac{1}{2}\left[1+{\rm erf}\left(\frac{\log_{10}(\rho_{\rm DM}/\rho_{\rm DM}^{\rm thre})}{\sigma_{\log\rho}}\right)\right],
\end{equation}
where erf is the error function, $\rho_{\rm DM}^{\rm thre}$ is a
threshold density dividing red and blue colors without scatter to
match the observed red fraction, and $\sigma_{\log\rho}$ denotes the
scatter in logarithmic scale of $\rho_{\rm DM}$. Fig.
\ref{fig:xil_rho_scatter} shows the comparison of observed quadrupole
components of the intermediate sample with the corresponding
measurements in the L.D. model by varying $\sigma_{\log\rho}$ as
$0.2$, $0.3$, and $0.4$. We change the parameter of $R_{\rm DM}$ to be
$600~$kpc to agree with the observed projected correlation functions
\citep{Masaki13a}.  Introducing an appropriate scatter improves the
agreement with the observations.  We find that the scatter value of
$\sigma_{\log\rho}$ around 0.3 has a best agreement in blue samples,
while $\sigma_{\log\rho}$ between 0.3 and 0.4 is best for red
samples. The chi-squared value for the $\sigma_{\log\rho}=0.3 (0.4)$
significantly decreases to be $3.62 (4.55)$ compared to the no-scatter case
of $6.11 (6.18)$ in Table \ref{tab:chi2}. The agreement in the age model is
still better than the L.D. model even when the scatter is
included. This is because the introduction of scatter in the form of
eq. [\ref{eq:scatter}] cannot reproduce the clustering of red and
blue galaxy samples simultaneously. This implies that the age model is
much better color assignment than the L.D. model and then the subhalo
age is a key ingredient to determine the color of galaxies.

\section{Summary and Conclusions}
\label{sec:summary}
We extend the SHAM approach to redshift space and test if the SHAM
explain the luminosity and color dependence of the redshift-space
clustering. We find that the simple subhalo abundance matching using
monotonic relation between the galaxy luminosity to the maximum
circular velocity qualitatively well reproduces the luminosity
dependence of SDSS galaxy clustering from $0.3~h^{-1}$Mpc to
$40~h^{-1}$Mpc. This indicates that the satellite motion inside
clusters is mainly determined with that of the host subhalos.  Our
results indicate that the SHAM method can be applied for RSD studies
and provides a promising way to construct mock samples of
redshift-space galaxy distribution. There is however a systematic
difference on subMpc scale and then the effect of baryonic physics may
be necessary to be included to achieve more precise theoretical
modeling for future galaxy surveys.

In addition to the luminosity dependence, we also apply two methods of
color abundance matching where the color is assigned by the age of
subhalos and the local DM density. We find that the color assignment
by the subhalo age much better agrees with the observations than by
the local DM density. The main reason why the local density model
fails to reproduce the observed clustering is that the fraction of red
subhalos in the model is too large inside the host halos. The
agreement improves by introducing a scatter in the relation between
color and local DM density, but still the color assignment based on
the subhalo age has better chi-squared values.  This suggests that the
subhalo age is a main driver of determining the color of galaxies and
a key ingredient to make an accurate mock galaxy samples with
different colors.

Our analysis using redshift-space clustering prefers the age model to
the local density model.  This is apparently inconsistent with the
galaxy-galaxy lensing analysis by \citet{Masaki13a}, which support the
local DM density model. This is because the redshift-space clustering
and lensing are sensitive to different aspects of the galaxy
clustering: the redshift-space clustering, in particular the FoG
effect, is sensitive to the satellite properties such as the satellite
profile and dynamics. On the other hand, galaxy-galaxy lensing probes
the averaged halo mass hosting all of the galaxies in a given
sample. This means that both of the color assignment models still need
to be improved.

In this analysis, we assume that the central galaxy sit on the halo
center and neglect the internal velocity relative to the host halo
bulk velocity.  Recent analysis using the Baryonic Oscillation
Spectroscopic Survey (BOSS) CMASS data shows that the central galaxies
may have $\sim 30$\% of the Virial velocity \citep{Guo15}. This
increases the velocity dispersion between central galaxies and
satellite galaxies at $\sim 9$\% ($\sim 0.3^2$). Baryon components
also affect the dynamics of satellite galaxies. The mean velocity of
satellites decrease by including baryonic components which prevents
the tidal disruption of slow subhalos by 10 percent level
\citep{Faltenbacher05,Wu13}. It may be also interesting to see how our
result changes using different algorithms of identifying subhalos. We
leave further detailed analysis as a future work.

\bigskip

\section*{Acknowledgments}
CH acknowledges support from a Grant-in-Aid for Scientific Research
from the Ministry of Education, Science, Sports, and Culture, Japan,
No. 24740160.

\bibliographystyle{mn2e}
\bibliography{mn-jour,refs}

\end{document}